# The average free volume model for the ionic and simple liquids

**Yang Yu*[a]**

[a] School of Physics and Optoelectronic Engineering, Nanjing University of Information Science & Technology, Nanjing 210044, Jiangsu, China; E-mail: yuyang5020@googlemail.com

In this work, the molar volume thermal expansion coefficient of 60 room temperature ionic liquids is compared with their van der Waals volume $V_w$. Regular correlation can be discerned between the two quantities. An average free volume model, that considers the particles as hard core with attractive force, is proposed to explain the correlation in this study. Some typical one atom liquids (molten metals and liquid noble gases) are introduced to verify this hypothesis. Good agreement between the theory prediction and experimental data can be obtained.

## Introduction

Comparing to the gas and the crystalline solid, the nature of the liquid is still unclear. There is no theory for the universally prediction of the properties of liquids from microscopic structure. On the other side, the liquid materials share many common phenomena. The Eötvos and Guggenheim empirical equations establish the relationship between surface tension, liquid density and critical temperature for the majority of liquids[1,2]. For the polymers, there is a correlation between crystalline volume $V_c$ and van der Waals volume $V_w$: $V_c \sim 1.435 V_w$[3]. For the ionic liquids, the correlation is $V_c$ (=$V_M$) ~ 1.410 $V_w$ ($V_M$ is molecular volume)[4]. The values 1.435 for polymers and 1.410 for ionic liquids are close to $2^{1/2}$, which corresponds to the minimum energy in Lennard-Jones (or 6-12) potential (at the position $r/\sigma = 2^{1/6}$, then the volume ratio is $(2^{1/6})^3 = 1.414$). [4-6]

The common works of liquids focus on the dynamics of particles and harsh repulsion within short range. And normally the attractive interaction is considered as introducing uniform background potential that provides the cohesive energy[7]. There are few studies working on the free space between molecules[8-12]. In this work, according to the relationship between thermal expansion coefficient and van der Waals volume for up to 60 ionic liquids, a new average free volume model – that is considering the atom or molecule as hard core with attractive force, each particle is surrounded by the average free volume – is established[12]. This work provides a new perspective to the liquid structure.To prove the validity of this new model, some typical one atom liquids are introduced for discussion and analysis.

## Results and discussion

From the experimental data[13,14], under the atmospheric pressure, the molar volume $V_{mol}$ displays linear correlation with temperature $T$ for all the ionic liquids, which can be presented by the linear function:

$$V_{mol}(T) = V_{E0} + C_1 T \qquad (1)$$

Here, $V_{E0}$ is the volume when extrapolate the molar volume in liquid state to absolute zero. The constant $C_1$ corresponds to the thermal expansion coefficient of the molar volume. Up to 60 ionic liquids are fit to equation (1), the experimental data are taken from the references as listed in table 1, which are collected in NIST website [13,15]. The fitting results of $V_{E0}$ and $C_1$ from equation 1 are compared with the van der Waals volume $V_w$ of the ionic liquids[4,16,17]. The van der Waals volume is the space occupied by a molecule, which is impenetrable to other molecules with normal thermal energies[6,18,19]. The fitting results and the van der Waals volume for each sample[4,16,17] are displayed in table 1. Obvious correlation between $V_{E0}$, $C_1$ and $V_w$ can be discerned from the figure 1.

Table1. The linear fitting results of equation 1 for the ionic liquids in this study. $V_w$: the van der Waals volume. $C_1$: thermal expansion coefficient of molar volume. $V_{E0}$: extrapolation of molar volume from liquid state to absolute zero.

| Name | Formula | $V_w^a$ cm$^3$/mol | $C_1$ cm$^3$/mol/K | $V_{E0}$ cm$^3$/mol | Ref[b] |
|---|---|---|---|---|---|
| 1-butyl-3-methylimidazolium methylsulfate | C$_9$H$_{18}$N$_2$O$_4$S | 139.1 | 0.1154 | 173.18 | 20 |
| 1-heptyl-3-methylimidazolium bis(trifluoromethylsulfonyl)imide | C$_{13}$H$_{21}$F$_6$N$_3$O$_4$S$_2$ | 209.5 | 0.2462 | 269.32 | 21 |
| 1-octyl-3-methylimidazolium bis[(trifluoromethyl)sulfonyl]imide | C$_{14}$H$_{23}$F$_6$N$_3$O$_4$S$_2$ | 219.7 | 0.2472 | 285.19 | 22 |
| 1-methyl-1-propylpyrrolidinium bis[(trifluoromethyl)sulfonyl]imide | C$_{10}$H$_{18}$F$_6$N$_2$O$_4$S$_2$ | 179 | 0.2017 | 224.73 | 23 |
| 1-ethyl-3-methylimidazolium tetrafluoroborate | C$_6$H$_{11}$BF$_4$N$_2$ | 99.3 | 0.0894 | 127.65 | 24 |
| 1-ethyl-3-methylimidazolium trifluoromethanesulfonate | C$_7$H$_{11}$F$_3$N$_2$O$_3$S | 115.1 | 0.1152 | 153.76 | 25 |
| 1-butyl-3-methylimidazolium nitrate[c] | C$_8$H$_{15}$N$_3$O$_3$ | 107 | 0.1021 | 144 | 26 |
| 1-decyl-3-methylimidazolium bis(trifluoromethylsulfonyl)imide | C$_{16}$H$_{23}$F$_6$N$_3$O$_4$S$_2$ | 240.2 | 0.2962 | 302.39 | 27 |
| 1-butyl-3-methylimidazolium octyl sulfate | C$_{16}$H$_{32}$N$_2$O$_4$S | 210.7 | 0.2037 | 266.97 | 28 |
| 1-butyl-3-methylimidazolium dicyanamide[c] | C$_{10}$H$_{15}$N$_5$ | 120.4 | 0.1162 | 158.44 | 29 |
| 1-butyl-1-methylpyrrolidinium bis[(trifluoromethyl)sulfonyl]imide | C$_{11}$H$_{20}$F$_6$N$_2$O$_4$S$_2$ | 191.8 | 0.1933 | 245.28 | 30 |
| 1-butyl-2,3-dimethylimidazolium hexafluorophosphate; | C$_9$H$_{17}$F$_6$N$_2$P | 142 | 0.1382 | 179.5 | 31 |
| 1-hexyl-3-methylimidazolium chloride | C$_{10}$H$_{19}$ClN$_2$ | 123 | 0.1082 | 162.72 | 32 |
| trihexyl(tetradecyl)phosphonium bis[(trifluoromethyl)sulfonyl]imide | C$_{34}$H$_{68}$F$_6$NO$_4$PS$_2$ | 440.1 | 0.4239 | 600.62 | 33 |
| 1-hexyl-3-methylimidazolium tetrafluoroborate | C$_{10}$H$_{19}$BF$_4$N$_2$ | 140.3 | 0.136 | 181.45 | 34 |
| 1-octyl-3-methylimidazolium tetrafluoroborate | C$_{12}$H$_{23}$BF$_4$N$_2$ | 160.7 | 0.1616 | 207.62 | 35 |
| 1-butyl-3-methylimidazolium 2-(2-methoxyethoxy)ethyl sulfate[c] | C$_{13}$H$_{26}$N$_2$O$_6$S | 181 | 0.1718 | 233.03 | 36 |
| 1-methyl-3-octylimidazolium 2-(2-methoxyethoxy)ethyl sulfate[c] | C$_{17}$H$_{34}$N$_2$O$_6$S | 222.7 | 0.2233 | 286.78 | 36 |
| pyridinium ethoxyethylsulfate[c] | C$_9$H$_{15}$NO$_5$S | 125.9 | 0.1141 | 160.62 | 37 |
| 1,3-dimethylimidazolium dimethylphosphate[c] | C$_7$H$_{15}$N$_2$O$_4$P | 108.8 | 0.0987 | 147.41 | 37 |
| 1-butyl-3-methylimidazolium dibutylphosphate[c] | C$_{16}$H$_{33}$N$_2$O$_4$P | 202.5 | 0.2104 | 269.12 | 38 |
| 1-ethylpyridinium bis(trifluoromethylsulfonyl)imide | C$_9$H$_{10}$F$_6$N$_2$O$_4$S$_2$ | 157.1 | 0.1623 | 204.44 | 37 |
| 1,3-dimethylimidazolium methoxyethylsulfate | C$_8$H$_{16}$N$_2$O$_5$S | 123.7 | 0.1051 | 160.68 | 37 |
| 1-butyl-3-methylimidazolium bis(perfluoroethylsulfonyl)imide | C$_{12}$H$_{15}$F$_{10}$N$_3$O$_4$S$_2$ | 214.3 | 0.234 | 273.34 | 39 |
| 1-butyl-3-methylimidazolium hexafluorophosphate | C$_8$H$_{15}$F$_6$N$_2$P | 131.8 | 0.1304 | 168.89 | 35 |
| 1-butylpyridinium bis[(trifluoromethyl)sulfonyl]imide | C$_{11}$H$_{14}$F$_6$N$_2$O$_4$S$_2$ | 177.5 | 0.1842 | 232.45 | 39 |
| 1-hexyl-3-methylimidazolium nitrate | C$_{10}$H$_{19}$N$_3$O$_3$ | 127.8 | 0.1167 | 170.52 | 40 |
| 1-butylpyridinium tetrafluoroborate | C$_9$H$_{14}$BF$_4$N | 118.5 | 0.1058 | 152.25 | 41 |
| 1-ethyl-3-methylimidazolium diethylphosphate | C$_{10}$H$_{21}$N$_2$O$_4$P | 140 | 0.1351 | 189.79 | 38 |
| 1-butyl-3-methylimidazolium tetrafluoroborate | C$_8$H$_{15}$BF$_4$N$_2$ | 119.8 | 0.1097 | 155.47 | 42 |
| butyltrimethylammonium bis(trifluoromethylsulfonyl)imide | C$_9$H$_{18}$F$_6$N$_2$O$_4$S$_2$ | 175.2 | 0.1837 | 230.02 | 43 |
| N,N-dimethyl-N-propyl-1-butanaminium bis(trifluoromethylsulfonyl)imide | C$_{11}$H$_{22}$F$_6$N$_2$O$_4$S$_2$ | 195.5 | 0.1859 | 259.12 | 33 |
| trihexyl(tetradecyl)phosphonium chloride[c] | C$_{32}$H$_{68}$ClP | 361 | 0.3585 | 481.27 | 33 |
| 1-pentyl-3-methylimidazolium bis(trifluoromethylsulfonyl)imide | C$_{11}$H$_{17}$F$_6$N$_3$O$_4$S$_2$ | 189 | 0.2116 | 245.44 | 44 |
| 1-butyl-3-methylimidazolium trifluoromethanesulfonate | C$_9$H$_{15}$F$_3$N$_2$O$_3$S | 135.6 | 0.1332 | 181.36 | 42 |
| 1,3-dimethylimidazolium bis[(trifluoromethyl)sulfonyl]imide | C$_7$H$_9$F$_6$N$_3$O$_4$S$_2$ | 148.1 | 0.1623 | 191.92 | 45 |
| 1-butyl-3-methylimidazolium tricyanomethane | C$_{12}$H$_{15}$N$_5$ | 137.7 | 0.1418 | 176.65 | 29 |
| 1-hexyl-3-methylimidazolium hexafluorophosphate | C$_{10}$H$_{19}$F$_6$N$_2$P | 152.3 | 0.1503 | 196.66 | 46 |
| 1-octyl-3-methylimidazolium hexafluorophosphate | $C_{12}H_{23}F_6N_2P$ | 172.7 | 0.1723 | 223.81 | 47 |
| hexyltrimethylammonium bis[(trifluoromethyl)sulfonyl]imide | $C_{11}H_{22}F_6N_2O_4S_2$ | 195.7 | 0.1956 | 265.34 | 33 |
| triethylhexylammonium bis[(trifluoromethyl)sulfonyl]imide | $C_{14}H_{28}F_6N_2O_4S_2$ | 226.5 | 0.2152 | 300.53 | 33 |
| 1-ethyl-3-methylimidazolium bis[(trifluoromethyl)sulfonyl]imide | $C_8H_{11}F_6N_3O_4S_2$ | 158.3 | 0.1732 | 206.16 | 43 |
| 1-ethyl-3-methylimidazolium ethyl sulfate | $C_8H_{16}N_2O_4S$ | 128.8 | 0.1061 | 159.41 | 48 |
| 1-butyl-4-methylpyridinium tetrafluoroborate; | $C_{10}H_{16}BF_4N$ | 128.5 | 0.1172 | 164.33 | 26 |
| Triisobutylmethylphosphonium tosylate[c] | $C_{20}H_{37}O_3PS$ | 236.8 | 0.2071 | 301.58 | 49 |
| 1-butyl-1-methylpyrrolidinium dicyanamide[c] | $C_{11}H_{20}N_4$ | 133.9 | 0.1031 | 155.99 | 26 |
| 1-ethylpyridinium ethylsulfate | $C_9H_{15}NO_4S$ | 127.6 | 0.0993 | 156.73 | 50 |
| 1-butyl-3-methylimidazolium thiocyanate[c] | $C_9H_{15}N_3S$ | 111.1 | 0.1031 | 153.68 | 51 |
| 1-ethyl-3-methylimidazolium dicyanamide[c] | $C_8H_{11}N_5$ | 99.6 | 0.0977 | 131.84 | 52 |
| 1-(2-hydroxyethyl)-3-methylimidazolium tetrafluoroborate[c] | $C_6H_{11}N_2OBF_4$ | 99.3 | 0.0836 | 138.69 | 53 |
| 1-butyl-3-methylimidazolium bis[(trifluoromethyl)sulfonyl]imide | $C_{10}H_{15}F_6N_3O_4S_2$ | 178.8 | 0.1974 | 232.82 | 43 |
| 1,3-diethylimidazolium bis[(trifluoromethyl)sulfonyl]imide | $C_9H_{13}F_6N_3O_4S_2$ | 168.6 | 0.2119 | 213.78 | 27 |
| 1-ethyl-2,3-dimethylimidazolium bis[(trifluoromethyl)sulfonyl]imide | $C_9H_{13}F_6N_3O_4S_2$ | 168.6 | 0.1693 | 221.5 | 54 |
| 1-octyl-3-methylimidazolium chloride | $C_{12}H_{23}ClN_2$ | 143.4 | 0.1338 | 188.73 | 20 |
| 1-ethyl-3-methylimidazolium bis[(pentafluoroethyl)sulfonyl]imide | $C_{10}H_{11}F_{10}N_3O_4S_2$ | 193.5 | 0.2143 | 244.53 | 55 |
| 1-hexyl-3-methylimidazolium bis[(trifluoromethyl)sulfonyl]imide | $C_{12}H_{19}F_6N_3O_4S_2$ | 199.3 | 0.2175 | 261.47 | 56 |
| 1,3-dimethylimidazolium methyl sulfate | $C_6H_{12}N_2O_4S$ | 108.4 | 0.0845 | 131.66 | 37 |
| 1-butyl-3-methylimidazolium perchlorate | $C_8H_{15}ClN_2O_4$ | 123.3 | 0.1079 | 158.33 | 57 |
| 1-butyl-3-methylimidazolium trifluoroacetate[c] | $C_{10}H_{15}F_3N_2O_2$ | 126.9 | 0.1288 | 168.84 | 58 |
| tetradecyl(trihexyl)phosphonium dicyanamide[c] | $C_{34}H_{68}N_3P$ | 386.2 | 0.3993 | 492.3 | 59 |

a. The values of $V_w$ are from Ue's and Machida's work [4, 16].
b. The density data are taken from the NIST website[13, 15], chosen from the references below.
c. When the $V_w$ for ions are not available from the Ue's and Machida's work, the value is calculated approximately by the method from the Zhao's work [17].

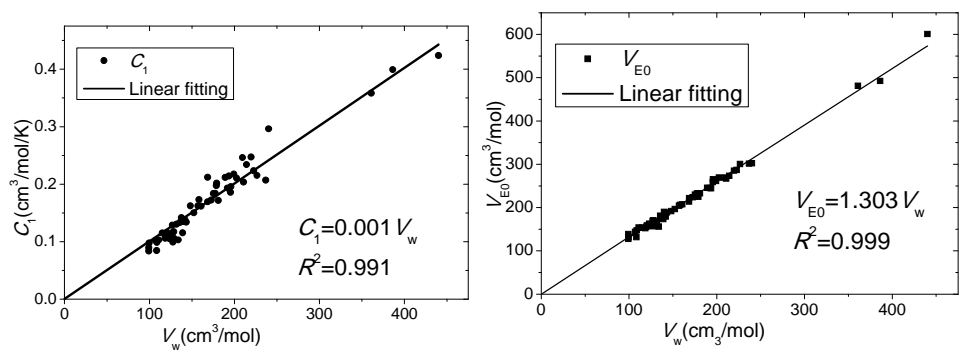

**Fig. 1** The correlation between $C_1$, $V_{E0}$ and $V_w$. The data are fitted by linear function fixing intercept at 0. $C_1$: thermal expansion coefficient of molar volume. $V_{E0}$: extrapolation of molar volume from liquid state to absolute zero. $V_w$: van der Waals volume.

The linearly fitting result $C_1$ and $V_{E0}$ of equation 1 are compared with $V_w$ as shown in figure 1. The $C_1$ shows a positive correlation with the $V_w$. The $V_{E0}$ is linearly increases with the $V_w$, and $V_{E0}/V_w$=1.303. This value is smaller than the crystalline volume ratio $V_c/V_w$= 1.410 for the ionic liquids[4], and it is close to the value of dense packing for the crystal 1.35 (reciprocal of 74.05%). When cancelling the intercept limitation in the fitting of $V_{E0}$ to $V_w$, the slope is 1.34. For the ionic liquids, considering the irregular shape of the ions, the deviation is acceptable.

**Proposed model:**

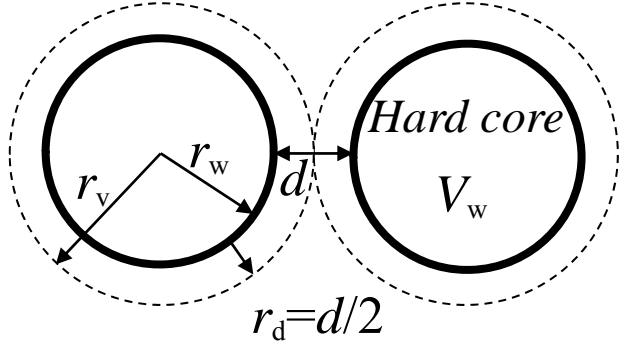

**Fig. 2** Schematic average free volume model for the liquids. The $r_w$ is the van der Waals radius. The $r_v$ is the radius of the larger sphere. The $d$ is distance between the two particles. The $r_d$ is half of the $d$.

The correlation between $C_1$, $V_{E0}$ and $V_w$ can be explained by a schematic average free volume model[12] displayed in figure 2. The molecules are simplified as spheres. Two particles are surrounded by the average free volume, and separated by the distance $d$, then the occupied volume for each particle $V$ is supposed to be the larger sphere volume (hard core volume $V_w$ adds the average free volume) plus interstitial volume $V_I$: $4\pi/3 \ast r_v^3 + V_I$, here $r_v = r_w + r_d$. So $V = 4\pi/3 \ast (r_w^3 + 3r_w^2 r_d + 3r_w r_d^2 + r_d^3) + V_I$. When temperature increases, $r_d$ increases, so $r_d$ is a function of $T$, $r_d(T)$. For simplicity, the first order approximation is applied to the formula, the second and third order of $r_d$ are omitted[12]. Then:

$$V(T) = V_w + V_I + 4\pi r_w^2 r_d(T) \qquad (2)$$

The comparison between equation (1) and (2) indicates the relationship:

$$V_{E0} = V_w + V_I \qquad (3)$$

Because the molar volume $V_{mol}$ linearly increases with temperature, so $r_d$ linearly changes with $T$, that is $r_d = C_2 T$, $C_2$ is constant, then:

$$C_1 = 4\pi r_w^2 C_2 = 4\pi C_2 [3V_w/(4\pi)]^{2/3} \qquad (4)$$

The $V_{E0}$ equals the van der Waals volume plus the interstitial volume, corresponding to the dense packing of crystal. And the thermal expansion coefficient $C_1$ has a positive correlation with van der Waals volume $V_w$.

Here, the average free volume $v_f$ is the free space averaged to each molecule, it is different from the local free (hole) volume $v_h$, which is the cavity in the real structure that can be seen from the positron annihilation lifetime spectroscopy (PALS) experiment. The hole volume $v_h$ is generated from coalescence of $v_f$ with statistics possibility because of the dynamic movement of particles[60-62].

From equation 4, the $C_1$ is not linearly increase with the $V_w$. The $C_2$ depends on the force and molecular movement between particles. Here the sphere is considered for simplification, when real particles are introduced, molecular structure and configuration should be accounted in. These are the reasons for the dispersion in the fitting of figure 1.

According to the average free volume model, the potential energy change can be presented in terms of surface tension as displayed in figure 3. The surface tension $\gamma$ expands the free volume, changes the sphere surface from $S_1$ to $S_2$, then the energy change with temperature in one degree is:

$$\Delta E_S = \gamma S_2 - \gamma S_1 = \gamma 4\pi (r_w + r_d + \Delta r_d)^2 - \gamma 4\pi (r_w + r_d)^2 \quad (5)$$
$$= \gamma 4\pi [2\Delta r_d (r_d + r_w) + \Delta r_d^2]$$

Since $\Delta r_d \ll r_w$, $r_d \ll r_w$ the terms $\Delta r_d^2$ and $r_d$ are omitted. $\Delta r_d = C_2 \Delta T$, $\Delta T = 1$ K. So:

$$\Delta E_S = 8\pi \gamma C_2 r_w \quad (6)$$

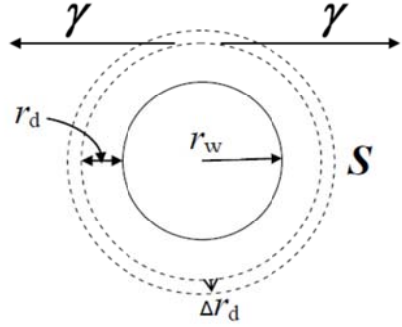

**Fig. 3** Energy change from the free volume expansion by the surface tension.

Combining equation (4) and (6) gives:

$$\Delta E_S = \frac{2\gamma C_1}{[3/(4\pi)]^{1/3} V_w^{1/3}} \quad (7)$$

The internal energy change under one degree of a monatomic fluid can be approximated as:

$$\Delta E = \frac{3}{2} R + \Delta E_S \quad (8)$$

According to the energy equipartition principle, the first term in equation 8 represents the kinetic energy, and the second term is the potential energy due to the expansion of the volume.

In order to check these theoretical predictions, the values $\Delta E_S$ and $\Delta E$ have been calculated for a number of metals and noble gases in liquid state at the melting point. These values are compared with the heat capacity of the liquids at melting point as shown in table 2.

Table 2. Comparison between the heat capacity and the energy change from the average free volume model.

| | $V_w$ [a] | $C_1$ [b] | $T_m$ [b] | $\gamma$ [b] | $\Delta E_S$ | $\Delta E$ | $C_p$ [c] |
|---|---|---|---|---|---|---|---|
| | cm$^3$/mol | cm$^3$/mol/K | K | N/m | J/mol/K | J/mol/K | J/mol/K |
| Metal | | | | | | | |
| Hg | 8.68 | 2.3*10$^{-3}$ | 234.29 | 0.489 | 14.90 | 27.37 | 28.47 |
| Al | 7.37 | 8.7*10$^{-4}$ | 933.45 | 1.050 | 12.78 | 25.25 | 29.26 |
| Na | 16.22 | 6.7*10$^{-3}$ | 370.95 | 0.200 | 14.41 | 26.88 | 31.81 |
| K | 29.48 | 1.4*10$^{-2}$ | 336.35 | 0.110 | 13.57 | 26.04 | 32.10 |
| Ca | 19.27 | 4.6*10$^{-3}$ | 1115 | 0.363 | 16.95 | 29.42 | 30.93 |
| Zn | 6.06 | 1.7*10$^{-3}$ | 692.68 | 0.789 | 20.03 | 32.50 | 31.35 |
| Fe | 5.04 | 9.4*10$^{-4}$ | 1811 | 1.881 | 28.07 | 40.54 | 43.89 |
| Ag | 7.53 | 1.1*10$^{-3}$ | 1234 | 0.926 | 14.15 | 26.62 | 30.51 |
| Cu | 5.29 | 6.9*10$^{-4}$ | 1357.77 | 1.320 | 14.23 | 26.70 | 31.35 |
| In | 11.74 | 1.9*10$^{-3}$ | 429.75 | 0.560 | 12.74 | 25.21 | 29.47 |
| Ga | 6.20 | 1.2*10$^{-3}$ | 302.91 | 0.724 | 12.87 | 25.34 | 27.80 |
| Noble gas | | | | | | | |
| Ne | 9.21 | 0.23 | 24.55 | 0.00566 | 16.91 | 29.38 | 43.08 |
| Ar | 16.75 | 0.13 | 83.8 | 0.01341 | 18.55 | 31.02 | 44.57 |
| Kr | 20.77 | 0.11 | 115.78 | 0.01641 | 17.88 | 30.35 | 43.33 |
| Xe | 25.40 | 0.10 | 161.36 | 0.01911 | 17.70 | 30.17 | 44.45 |

a. The $V_w$ of metal is calculated from metallic radii, assuming that the atoms are spherical. For the noble gas, the van der Waals radii is chosen to calculate the $V_w$.[63]
b. The values of density to calculate the thermal expansion coefficient of molar volume $C_1$, the melting temperature $T_m$, the surface tensiton $\gamma$ are from the book[64].
c. The heat capacity $C_p$ for the metal is obtained from reference[65, 66]. For the noble gas, the $C_p$ is obtained from the NIST database[67].

Considering the simplicity of the model, the theoretical prediction fits the experimental result very well without any adjustable parameter. From the result in table 2, the deviation between the calculated result and the experimental data of the noble gas is larger than the molten metal. One possible reason is, in the average free volume model, the pressure is not accounted in, for the metal, this force can be neglected, but the pressure added is comparable with small interaction force between the atoms for the noble gas, more energy is needed to overcome the additional force. The detailed work is needed to improve the model.

## Conclusions

According to the correlations between the van der Waals volume and the molar volume thermal expansion coefficient as well as extrapolate volume at absolute zero, an average free volume model is proposed. The volume can be estimated only with the information of van der waals volume for the ionic liquids: $V_{mol}=0.001V_wT+1.303V_w$.

The diverse range of fluids from ionic liquids to molten metals and liquid noble gases suggests the validity of this model.

This average free volume model is more a technique method than corresponding to a real structure, since from the result and discussion of our previous experiment PALS work, the average $v_f$ will coalesce to larger holes[12, 60-62].

The first order approximation is applied when calculate the volume and potential energy change.

It is questionable to use $V_w$ as hard core volume. Precise work is needed for this part.

Because of the simplicity of the model, the extension investigation is needed, such as the external force caused by pressure, the multi-atom molecules other than one atom substances, the shape factor and so on.

## Acknowledgements

Dr. Yang Yu acknowledges the financial support from the National Natural Science Foundation of China (Grant No.: 11247220) and the Science Research Startup Foundation from the Nanjing University of Information Science & Technology (Grant No: S8112078001). This work is also supported by the Natural Science Foundation of Jiangsu Province (No. BK20131428 )